%% file: t2hkdune_draft.tex
\documentclass[aps,prd,superscriptaddress,nofootinbib,tighten,preprint]{revtex4}
\include{header}
\usepackage{amsfonts}
\usepackage{amssymb}
\usepackage{amsmath}
 \pdfoutput=1
\usepackage{graphicx}
\usepackage{hyperref}
\usepackage{placeins}
\usepackage{gensymb}
\usepackage{xcolor}
\def\slc#1{\setbox0=\hbox{$#1$}           
    \dimen0=\wd0                                 
    \setbox1=\hbox{/} \dimen1=\wd1               
    \ifdim\dimen0>\dimen1                        
       \rlap{\hbox to \dimen0{\hfil/\hfil}}      
       #1      =                                  
    \else                                        
       \rlap{\hbox to \dimen1{\hfil$#1$\hfil}}   
       /                                         
    \fi}

\newcommand{\beq}{\begin{equation}}
\newcommand{\eeq}{\end{equation}}
\newcommand{\beqa}{\begin{eqnarray}}
\newcommand{\eeqa}{\end{eqnarray}}

\newcommand{\ty}{{\theta_{13}}}
\newcommand{\tz}{{\theta_{23}}}

\begin{document}

\title{Imprints of a light Sterile Neutrino at DUNE, T2HK and T2HKK }

\author{Sandhya Choubey}
\email{sandhya@hri.res.in}
\affiliation{Harish-Chandra Research Institute, Chhatnag Road, Jhunsi, Allahabad 211 019, India}
\affiliation{Department of Theoretical Physics, School of
Engineering Sciences, KTH Royal Institute of Technology, AlbaNova
University Center, 106 91 Stockholm, Sweden}
\affiliation{Homi Bhabha National Institute, Training School Complex, Anushakti Nagar,
     Mumbai 400085, India
}

\author{Debajyoti Dutta}
\email{debajyotidutta@hri.res.in}
\affiliation{Harish-Chandra Research Institute, Chhatnag Road, Jhunsi, Allahabad 211 019, India}
\affiliation{Homi Bhabha National Institute, Training School Complex, Anushakti Nagar,
     Mumbai 400085, India
}

\author{Dipyaman Pramanik}
\email{dipyamanpramanik@hri.res.in}
\affiliation{Harish-Chandra Research Institute, Chhatnag Road, Jhunsi, Allahabad 211 019, India}
\affiliation{Homi Bhabha National Institute, Training School Complex, Anushakti Nagar,
     Mumbai 400085, India
}

\begin{abstract}

We evaluate the impact of sterile neutrino oscillations 
in the so-called 3+1 scenario on the proposed long baseline experiment in 
USA and Japan.  
There are two proposals for 
the Japan experiment which are called T2HK and T2HKK.
We show the impact of sterile neutrino oscillation parameters 
on the expected sensitivity of T2HK and T2HKK 
to mass hierarchy, CP violation and octant of $\theta_{23}$ and compare it against 
that expected in the case of standard oscillations. We add the expected 
ten years data from DUNE and 
present the combined 
expected sensitivity of T2HKK+DUNE to the oscillation parameters.
We do a full marginalisation over the 
relevant parameter space and show the effect of the magnitude of the true 
sterile mixing angles on the physics reach of these experiments. We show that 
if one assumes that the source of CP violation is the standard CP phase alone in 
the test case, then 
it appears that the expected CP violation sensitivity decreases due to sterile neutrinos. 
However, 
if we give up this assumption, then the CP sensitivity could go in 
either direction. The impact on expected octant of $\theta_{23}$ and 
mass hierarchy sensitivity is shown to 
depend on the magnitude of the sterile mixing angles in a nontrivial way.

\end{abstract}
\maketitle
\section{Introduction}

The field of neutrino oscillations is on the verge of its final leap forward. 
The three-generation paradigm has been firmly established and the leading oscillation parameters are
determined to very high precision. 
The last of the neutrino oscillation parameters 
are expected to be measured with unprecedented precision in the next generation long baseline experiments. 
In particular, the questions regarding CP violation in the lepton sector, the order of the neutrino masses and the octant of the 
mixing angle $\theta_{23}$ are expected to be answered within the next decade. The most promising experiments that 
the neutrino community is looking forward to in this regard are the Deep Underground Neutrino Experiment (DUNE) in USA 
\cite{Acciarri:2016ooe,Strait:2016mof,Acciarri:2015uup,Acciarri:2016crz}
and the Tokai to Hyper-Kamiokande (T2HK) experiment in Japan \cite{Abe:2015zbg}. 
The question on neutrino mass hierarchy should be the first to be resolved with 
the expected sensitivity reach of DUNE being more than $5\sigma$. The detector for the T2HK experiment 
in Japan is proposed to be placed at only 295 km and hence with a narrow-band beam peaked at 
about 0.6 GeV, the expected matter effects in T2HK is low and its mass hierarchy sensitivity rather poor. 
To circumvent this problem there is a proposal to shift one module and hence half of the HK detector to 
a location in Korea at a distance of 1100 km. This proposal is referred to as T2HKK (Tokai to Hyper-Kamiokande Korea).\footnote {In this work we will refer to the first module of the megaton HK detector as JD (Japan Detector) while 
the second module will be called KD (Korea Detector).}
This brings a more than $3\sigma$ discovery potential for a large fraction of the true CP phase $\delta_{13}$(true).
While DUNE is expected to discover CP violation at the $3 \sigma$ level for 
75\% of the true $\delta_{13}$ (with 1320 kt. MW. year) \cite{Acciarri:2015uup}, T2HK is expected to discover it at the $5 \sigma$ level for 
55-60\% of the true $\delta_{13}$ \cite{Abe:2016ero}. 
The question on the true octant of $\theta_{23}$ should be settled at the 
$5 \sigma$ level for $\sin^2\theta_{23}$(true)$ < 0.43$ and $\sin^2\theta_{23}$(true)$ > 0.59$ with 
data at DUNE and for $\sin^2\theta_{23}$(true)$ < 0.46$ and $\sin^2\theta_{23}$(true)$ > 0.56$ with 
data at T2HK \cite{Abe:2015zbg}.

Even though neutrino oscillations within the three-generation framework is well established, 
anomalous hints of neutrino oscillations at a higher frequency corresponding to a mass squared 
difference $\Delta m^2 \sim 1$ eV$^2$ have been around since a long time \cite{Abazajian:2012ys}. The first such 
claim came from the LSND experiment \cite{Athanassopoulos:1995iw,Aguilar:2001ty} 
in Los Alamos, USA, where a $\bar\nu_\mu$ beam 
was sent to a detector and the observations showed a $3.8\sigma$ excess in the electrons (positrons)  
which could be explained in terms of $\bar\nu_\mu \to \bar\nu_e$ oscillations driven by $\Delta m^2 \sim 1$ eV$^2$. 
This oscillation claim was tested by the KARMEN \cite{Gemmeke:1990ix}
and MiniBooNE \cite{AguilarArevalo:2007it,Aguilar-Arevalo:2013pmq,AguilarArevalo:2010wv} experiments. 
While KARMEN data did not show any evidence for $\nu_\mu \to \nu_e$ oscillations, it was unable to 
rule out the entire parameter space allowed by LSND. More recently, the MiniBooNE experiment 
ran in both the neutrino as well as antineutrino mode. The MiniBooNE collaboration 
concluded that while they did not see any electron excess in their neutrino event sample 
consistent with LSND, their antineutrino data shows an excess confirming the oscillations 
seen by LSND. MiniBooNE reported an excess of electrons/positrons at low energies in both 
neutrino as well as antineutrino modes, however these excess events are not coming from 
neutrino flavor oscillations. 

More recently, anomalies in the reactor antineutrino flux measurements \cite{Mention:2011rk,Mueller:2011nm,Huber:2011wv} 
and the gallium experiments calibration data \cite{Abdurashitov:2005tb,Bahcall:1994bq,Giunti:2006bj,Giunti:2010wz,Giunti:2010zu}
have also hinted at  the possibility of $\nu_e$ (or $\bar\nu_e$) oscillations 
with $\Delta m^2 \sim 1$ eV$^2$. Since it is not possible to accommodate this extra mass 
squared difference within the three-generation framework, one needs to add at least one 
additional light neutrino species and since the Z decay width constrains the number of 
light neutrinos coupled to  the Z-boson to be 2.9840$\pm$0.0082 \cite{ALEPH:2005ab}, 
the additional neutrino(s) should not have any standard model gauge interactions. 
These neutrinos are hence called sterile for this reason. With one additional sterile 
neutrino, one could have a neutrino mass spectrum of the so-called 2+2 or 3+1 type \cite{Goswami:1995yq}, 
where the former has two lighter neutrinos separated by two heavier ones by $\Delta m^2 \sim 1$ eV$^2$, 
while the latter has one heavier state separated by three lighter ones by $\Delta m^2 \sim 1$ eV$^2$.
The 2+2 mass spectrum predicts oscillations involving the sterile neutrino in either the solar or the atmospheric neutrino sectors. 
Since neither of the two data sets allow this, the 2+2 spectrum 
is heavily disfavored by the solar neutrino and atmospheric neutrino observations, 
as well as from bounds on the sum of neutrino masses from cosmology \cite{Lattanzi:2016rre}.
The 3+1 spectrum on the other hand predicts low admixture of the sterile neutrinos in solar and atmospheric 
neutrino oscillations and hence survives. It also has only one heavy mass state instead of two and hence has 
less issues with the bounds from cosmology. Nonetheless, the global analysis of 
all relevant data sets that constrain the 3+1 parameter space returns a goodness of fit 
of only 19\% \cite{Kopp:2013vaa}. 
The main problem with the 3+1 scenario is the tension between the appearance data 
in LSND and MiniBooNE and the disappearance data of $\nu_\mu$ at 
at experiments like CDHS and MINOS.

It has been pointed out \cite{Klop:2014ima,Gandhi:2015xza}
that the sensitivity of long baseline experiments to standard neutrino 
oscillation parameters such as neutrino mass hierarchy, CP violation and octant of $\theta_{23}$ gets 
compromised in the presence of sterile neutrino(s). As we will discuss, this happens because 
even though the oscillations due to the $\Delta m^2 \sim $ 1 eV$^2$ frequency 
average out at the long baselines, the impact of the additional mixing angles and 
phases due to the additional sterile neutrinos stay in the relevant oscillation probabilities.  
The expected sensitivity of the DUNE near detector to the sterile neutrino parameters themselves 
was studied in \cite{Choubey:2016fpi} while the corresponding sensitivity of the DUNE far detector 
to these parameters were presented in \cite{Berryman:2015nua}. The expected sensitivity of T2HK  to the sterile neutrino parameters has been studied recently in \cite{Kelly:2017kch}.
The impact of sterile neutrinos 
on the sensitivity to standard oscillation parameters in T2K and NO$\nu$A was discussed in \cite{Klop:2014ima,Agarwalla:2016mrc,Ghosh:2017atj}, 
and in DUNE in \cite{Berryman:2015nua,Gandhi:2015xza,Agarwalla:2016xxa,Agarwalla:2016xlg,Dutta:2016glq}.  
In particular, the expected change to 
standard neutrino parameter sensitivity of DUNE for the 3+1 scenario has been given in \cite{Dutta:2016glq,Berryman:2015nua}, 
taking into account the marginalisation of the sterile neutrino as well as the relevant standard oscillation parameters.

In this work we focus on two main purposes. Firstly, we 
consider the impact of sterile neutrino mixing in the 3+1 scenario on the 
physics reach of the T2HK and T2HKK 
set-ups. We calculate the expected sensitivity to 
neutrino mass hierarchy, CP violation and octant of $\theta_{23}$
both in the absence (referred to as the 3+0 case) as well as presence 
(referred to as the 3+1 case) of a sterile neutrino. 
Secondly, we present combined analysis of the T2HKK set-up with DUNE to check for 
possible synergies between the two experiments. We present the 
expected sensitivity of the T2HK and T2HKK set-ups to the oscillation
parameters in both 3+0 and 3+1 scenarios. 

The paper is organised as follows. We will discuss the oscillation probabilities in 
the 3+1 case in section II. The simulation procedure will be given in section III, 
while the experimental configurations considered in our analyses 
will be specified in section IV. We will discuss our results and present the expected 
sensitivities of the T2HK, T2HKK and T2HKK+DUNE configurations in section V. 
Finally, we will conclude in section VI. 

\section{Sterile Neutrino Hypothesis}
In order to explain the short baseline anomalies \cite{Abazajian:2012ys}, one or more extra neutrino states of mass squared difference $\sim$ 1 eV$^{2}$ are proposed. But the LEP data suggests that these extra neutrinos must be singlet under standard model gauge group. Thus they are called sterile neutrinos. For the 3+1 case, the full neutrino mixing matrix will have six mixing angles, three phases and three mass-squared differences. This mixing matrix can be parametrized in the following way:
\begin{equation}
\label{para}
U^{3+1}_{\rm PMNS} = O(\theta_{34},\delta_{34})O(\theta_{24},\delta_{24})R(\theta_{14})R(\theta_{23})O(\theta_{13},\delta_{13})R(\theta_{12})
\end{equation}
Here, $O(\theta_{ij},\delta_{ij})$ are $4\times4$ orthogonal matrices with associated phase $\delta_{ij}$ in the $ij$ sector, and $R(\theta_{ij})$ are the rotation matrices in the $ij$ sector. Since $\Delta m^{2}_{41} \sim 1$ eV$^2$, the oscillations driven by this mass parameter would be averaged out at the far detector of T2HK, T2HKK as well as DUNE. 
We can see that in this parametrisation, the  vacuum $\numu\rightarrow\nue$ oscillation probability is independent of the $\theta_{34}$ mixing angle and the associated phase. However, in the long baseline experiments like DUNE or T2HKK, there is enough matter effects so that this independence is lost and the neutrino appearance probabilities do exhibit a significant dependence on them at these long baseline experiments \cite{Gandhi:2015xza}. 

Again, $\nu_e$ appearance probability in vacuum contains some terms which are dependent on sine and cosine of $\delta_{13}$, $\delta_{24}$ and $(\delta_{13}+\delta_{24})$. 
These interference terms connecting the mixing angles from both 3+0 and 3+1 model, can not only enhance the amplitude of the probability but also affect the sensitivity of long baseline experiments \cite{Gandhi:2015xza,Dutta:2016glq}. So keeping this in mind, here we study the effect of this new mass eigenstate at the far detector of these experiments.

\section{Simulation procedure}

In our analysis we have used GLoBES (Global Long Baseline Experiment Simulator) \citep{Huber:2004ka,Huber:2007ji} to perform all our computations, along with additional sterile neutrino codes \cite{Kopp:2006wp,Kopp:2007ne}. In all the results presented in this paper, we do a full four-generation analysis in the 3+1 framework and calculate the oscillation probabilities in matter using exact numerical codes. Through out this analysis we choose the true values of the standard neutrino oscillation parameters as follows \cite{Esteban:2016qun}:
\be
 \theta_{12}=33.5\degree,~\theta_{13}=8.5\degree,~\ms=7.5\times10^{-5}~{\rm eV}^{2},~\ma=2.52\times10^{-3}~{\rm eV}^{2} 
 \ee
We choose true value of $\theta_{23} = 45.0\degree$ for CPV and mass hierarchy studies, whereas for the octant sensitivity studies 
we generate data at $\theta_{23}=40.3\degree$ (we call this case Lower Octant (LO)) 
and $49.7\degree$ (we call this case Higher Octant (HO)). The true mass hierarchy is always taken as normal hierarchy. 
The true $\delta_{13}$ is varied in its full range [$-\pi,\pi$]. 
Since there are twelve parameters in the 3+1 scenario, it is impossible to marginalise over all parameters. Therefore, we check 
the impact of the relevant parameters on each sensitivity study performed in this work and marginalise the resultant 
$\chi^2$ only over the ones that bring any significant change. We will state clearly our marginalisation procedure for every 
result shown in this paper. Where ever relevant, the marginalisation is performed for 
$\theta_{13}$, $\theta_{23}$ and $\ma$ in the ranges are $[7.99^0, 8.91^0]$, $[38.4^0, 53.0^0]$ and [$2.40, 2.64$]$\times10^{-3}$ eV$^2$ for NH \& [$-2.64, -2.40$]$\times10^{-3}$ eV$^2$ for IH, respectively.

The issue regarding sterile neutrino mass squared difference and mixing angles is far from settled. While LSND \cite{Athanassopoulos:1995iw,Aguilar:2001ty}, MiniBooNE \cite{AguilarArevalo:2007it,Aguilar-Arevalo:2013pmq,AguilarArevalo:2010wv}, 
gallium \cite{Abdurashitov:2005tb,Bahcall:1994bq,Giunti:2006bj,Giunti:2010wz,Giunti:2010zu} and reactor anomaly \cite{Mention:2011rk,Mueller:2011nm,Huber:2011wv} require sterile neutrinos mixed with the active ones, the Bugey-3 \cite{Declais:1994su}, 
Daya Bay \cite{ An:2014bik,Adamson:2016jku}, NEOS \cite{Ko:2016owz}, IceCube  \cite{TheIceCube:2016oqi}, MINOS \cite{Adamson:2011ku} and NO$\nu$A 
\cite{Adamson:2017zcg} experiments are consistent with no active-sterile oscillations and put severe upper limits on the mixing angles 
$\theta_{14}$ \cite{Bora:2012pi, Declais:1994su, An:2014bik, Adamson:2016jku}, $\theta_{24}$ \cite{TheIceCube:2016oqi} and 
$\theta_{34}$  \cite{Adamson:2011ku} (see also \cite{Adamson:2017zcg}). The phases are completely unconstrained at the moment. 
Since there is a certain degree of uncertainty regarding the active-sterile mixing angles, we 
perform our analysis for two sets of benchmark true values:
\be
(\theta_{14}, \theta_{24}, \theta_{34}) &=& (6.5^{o}, 3.5^{o}, 12.5^{o}) \,,\\
(\theta_{14}, \theta_{24}, \theta_{34}) &=& (12^{o}, 7^{o}, 25^{o})\,,
\ee
where the former is the set of smaller values of the mixing angles, while the latter are at the higher values. The marginalisation over 
these mixing angles are done in the range $\theta_{14}\leq13\degree$ \cite{An:2014bik}, $\theta_{24}\leq7\degree$  \cite{TheIceCube:2016oqi} 
and $\theta_{34}\leq26\degree$ \cite{Adamson:2011ku}. The phases are taken over their full range [$-\pi,\pi$]. The mass squared difference was 
taken fixed at $\Delta m^2_{41}=1$ eV$^2$. The true value of $\Delta m^2_{41}$ (as well marginalisation over it) has absolutely no impact 
on our results since the oscillations driven by $\Delta m^2_{41}$ get effectively averaged as long as they are significantly higher than $10^{-3}$ eV$^2$. 
\footnote{Recently, in \cite{Gariazzo:2017fdh}, the authors have put strong constraints on the sterile mixing angles from global analysis. The upper limit, used in this work for $\theta_{34}$ is far outside the range specified in \cite{Gariazzo:2017fdh}. But we have explicitly checked our results with the new constrain on $\theta_{34}$ and found that the results reported in this work will not change by 5$\%$.}

\section{Experimental setup}

\subsection{T2HK}

The Hyper-Kamiokande (HK) \cite{Abe:2015zbg} is the upgradation of the Super-Kamiokande (SK) \cite{Fukuda:1998mi} program in Japan, where the detector mass is proposed to be increased by about twenty times the fiducial mass of SK. The main purpose of this huge detector is to pursue neutrino and proton decay study. The proposed HK consist of two 187 kt water Cherenkov detector modules, near the current SK site, 295 km away and 2.5$\degree$ off-axis (OA) from the J-PARC beam which is currently being used by the T2K experiment \cite{Abe:2016ero}. The long baseline experiment T2HK has similar physics goals as DUNE, but being a narrow beam it can be complimentary to the DUNE experiment. In this article we will alternately call this set-up T2HK or JD$\times$2 which stands for the two modules of Japan Detector. 

For our analysis we have taken a beam power of 1.3 MW and 2.5$\degree$ off-axis flux. We consider 
a baseline of 295 km and total fiducial mass of 374 kt (assuming two tanks each of 187 kt). Total run time of 10 years is divided between neutrino and anti-neutrino mode in 1:3 ratio. For the energy resolution, we assume a Gaussian function of width 15$\%/\sqrt{E}$. We have matched the number of events given in TABLE $\rm II$ and TABLE $\rm III$ of \cite{Abe:2016ero}. The signal normalization error in $\nue$ ($\bar{\nue}$) appearance and  $\numu$ ($\bar{\numu}$) disappearance channel are  3.2\% (3.6\%) and 3.9\% (3.6\%), respectively. The background and energy calibration error in all the channels are 10\% and 5\%, respectively. 
\subsection{T2HKK}
In \cite{Abe:2016ero}, the collaboration has discussed the possibility of shifting the second detector module of the HK setup to Korea. Thus, T2HKK experiment will have two sets of detectors; one at the HK site at a baseline of 295 km and the other one at Korea at a baseline of $\sim$1100 km. The second detector module will be a 187 kt water Cherenkov detector identical to the HK detector module in Kamioka. 
The second oscillation maximum takes place near $E_{\nu} = 0.6$ GeV at the $\sim$1100 km baseline. 
If the detector is placed at 2.5$\degree$ off-axis, the peak energy of the narrow beam coincides with 
the second oscillation maximum at 0.6 GeV. We will refer to this set-up alternately as T2HKK or JD+KD which impleis one detector at Japan while the other at Korea. The background and systematic uncertainties are kept same as T2HK.

\subsection{DUNE}
DUNE (Deep Underground Neutrino Experiment) \cite{Acciarri:2016crz,Acciarri:2016ooe,Acciarri:2015uup,Strait:2016mof}  
is a future long baseline experiment which is supposed to address all the three important issues in the neutrino sector -- determination of neutrino mass hierarchy, leptonic CP violation, and the octant of $\theta_{23}$. The neutrino beam source will be at Fermilab in Chicago, IL and the far detector will be at Sanford Underground Research Facility (SURF) in South Dakota, at a distance of 1300 km from the beam source. The accelerator facility at Fermilab will give a proton beam of energy 80-120 GeV at 1.2-2.4 MW. The proton beam will eventually give a wide neutrino beam of energy range 0.5-8.0 GeV which will be detected at the far site. The far site will consist of 4 identical detector of about 10 kt each. The  detector will be a liquid Argon time projection chamber (LArTPC).

In our analysis, we have assumed that a Liquid Argon (LAr) detector of fiducial mass 34 kt will be installed at a baseline of 1300 km 
and the 120 GeV-1.2 MW proton beam will deliver 10$^{20}$ protons on target per year. We assume total run time of 10 years which is equally divided between neutrinos and anti-neutrinos and this will give a total exposure of $35\times10^{20}$ kt-POT-yr. The energy resolution considered here for $\mu$ and $e$ are $20\%/\sqrt{E}$ and $15\%/\sqrt{E}$, respectively. We have combined both $\nue$ appearance and $\numu$ disappearance channels, both in neutrino and anti-neutrino mode. The signal efficiency is taken to be 85\%. All other backgrounds are taken from \citep{Acciarri:2015uup}. In the $\nu$ ($\bar{\nu}$) mode, the signal normalisation error is 2\% (5\%), the background normalisation error is 10\% (10\%) and the energy calibration error is 5\% (5\%). The choice of systematics is conservative compared to the projected systematics in \citep{Acciarri:2015uup}.

In Table~\ref{Table 2}, we give the total number of expected events for all channels in all the three experiments for both the 3+0 and the 3+1 scenarios. The value of all the phases have been taken as zero, while $\theta_{23}=45\degree$. The values of the other oscillation parameters used have been given before in section III. For both JD and KD we divide the total run time of 10 years in the ratio of 1:3 between $\nu$ and $\bar{\nu}$, while in DUNE the ratio is 1:1. The events in the 3+1 case is generated for the smaller set of mixing angles given in section III. 

 \begin{table}[!h]
\begin{center}
\begin{tabular}{|c|c|c|c|c|}
\hline 
Experiments & Channels & 3+0   & 3+1   \tabularnewline 
\hline 
JD & $\nu_e (\bar{\nu_e})$ app & 1839 (1715)  & 1865 (1738)\tabularnewline
 
 & $\nu_{\mu} (\bar{\nu_{\mu} })$ disapp & 10478 (14435)   & 10390 (14345) \tabularnewline
\hline 
KD &$\nu_e (\bar{\nu_e})$ app & 132 (165) & 134 (169)   \tabularnewline
 
 & $\nu_{\mu} (\bar{\nu_{\mu} })$ disapp & 1373 (1477)& 1357 (1437)\tabularnewline
\hline 
DUNE &$\nu_e (\bar{\nu_e})$ app & 1101 (284) & 1146 (296)  \tabularnewline
 
 & $\nu_{\mu} (\bar{\nu_{\mu} })$ disapp & 7419 (5086)   & 7326 (5054) \tabularnewline
\hline 
\end{tabular}
\caption{Total number of events in JD, KD and DUNE in both the scenarios. In the 3+0 case, we take $\delta_{13}=0^0$ while in the 3+1 case, all the three phases are taken as zero. For both JD and KD we divide the total run time of 10 years in the ratio of 1:3 between $\nu$ and $\bar{\nu}$, while in DUNE the ratio is 1:1. The events in the 3+1 case is generated for the smaller set of mixing angles given in section III. The take $\theta_{23}=45\degree$ for all cases. The values of the other oscillation parameters taken are given in section III. }
\label{Table 2}

\par\end{center}
\end{table}

\section{Results}
\subsection{Numerical probabilities}

\begin{figure}
\includegraphics[scale=0.5]{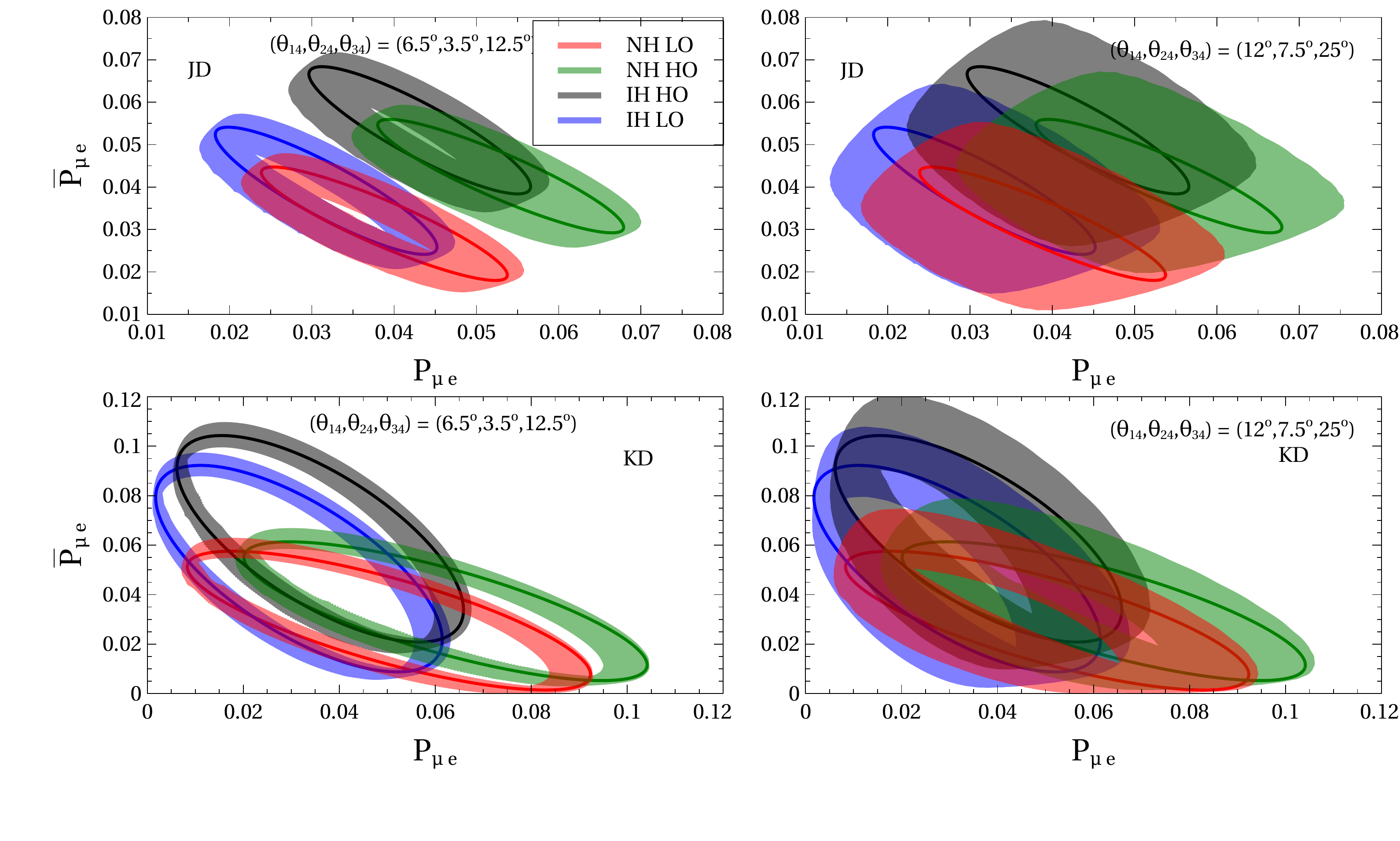}
\caption{$P_{\mu e}$ vs $\bar{P}_{\mu e}$ bi-probability plots for JD and KD. The first row shows the plots for JD (295 km and $E=0.5$ GeV ) and the second row shows the plots for KD (1100 km and $E=0.66$ GeV). The left column is for lower set of sterile mixing angles and the right column is for higher set of sterile mixing angles. The shaded regions show the variation of the bi-probability when the sterile phases $\delta_{24}$ and $\delta_{34}$ are varied  within [-$\pi$,$\pi$]. The solid curves describe the 3+0 case. The different colors correspond to different combinations of mass hierarchy and octant of $\theta_{23}$. }
\label{biprob}
\end{figure}

In this section we present some of our results at the probability level in order to understand the physics potential of these experiments. The Fig.~\ref{biprob} gives the bi-probability plots for JD (the Japan detector at 295 km) at its first oscillation maximum at 0.5 GeV and KD (the Korean detector at 1100 km) at its second oscillation maximum at 0.66 GeV. For each case we have considered all the four combinations of octant (Higher Octant (HO) or Lower Octant (LO)) and mass hierarchy (Normal Hierarchy (NH) or Inverted Hierarchy (IH)) {\it i.e.}, NH-LO (red), IH-LO (blue), NH-HO (green) and IH-HO (black). As stated before, for LO we have chosen $\theta_{23}=40.3\degree$ as the benchmark value and for HO we take $\theta_{23}=49.7\degree$. For NH we have chosen $|\Delta m^{2}_{31}|=2.457\times10^{-3}$eV$^{2}$ and for IH we have chosen $|\Delta m^{2}_{31}|=2.449\times10^{-3}$eV$^{2}$. 
The effect of the variation in the bi-probability plots due to the sterile sector phases $\delta_{24}$ and $\delta_{34}$ varying in their 
[$-\pi,\pi$] range is shown by the shaded regions. For comparison we also show the bi-probability ellipse for the 
standard 3+0 case by the solid curves of the same color.

The top-left panel shows that for JD there is a strong overlap for NH and IH bi-probability curves even for the 3+0 case and 
for both LO and HO. This reflects the fact that JD has a smaller baseline and energy and hence small matter effects. 
As a result, the sensitivity of JD to CP violation is expected to suffer from mass hierarchy degeneracy. 
As we switch on the sterile neutrino mixing, the sharp bi-probability lines start to spread and there is a whole band 
of bi-probabilities coming from the fact that now in addition to $\delta_{13}$ there are 2 additional phases $\delta_{24}$ and 
$\delta_{34}$. The bi-probability bands for the different colors representing different cases start to overlap more with 
each other. The overlap increases significantly as we increase the sterile mixing angles, shown in the 
top-right panel, owing to the fact that the 
effect of the phases $\delta_{24}$ and $\delta_{34}$ increase. 
The corresponding bi-probability results for the KD baseline and peak energy of 0.66 GeV is shown in the lower panels.
As for JD, the bi-probability curves for 3+0 overlap between the different hierarchies even for KD, but the overlap is less. 
As we switch on the sterile mixing angle (bottom-left panel) the bi-probability lines blur into a band, though the spread in 
the case of KD appears to be slightly less. As the sterile mixing angles increase (bottom-right panel) the 
blurring increases showing the increased effect of the phases associated with the sterile mixing angles. We have checked that 
the main contribution to the blurring of the bi-probability plot comes from the phase $\delta_{24}$. The effect of $\delta_{34}$ 
is subdominant, although non-zero. 


\begin{figure}
\includegraphics[scale=0.75]{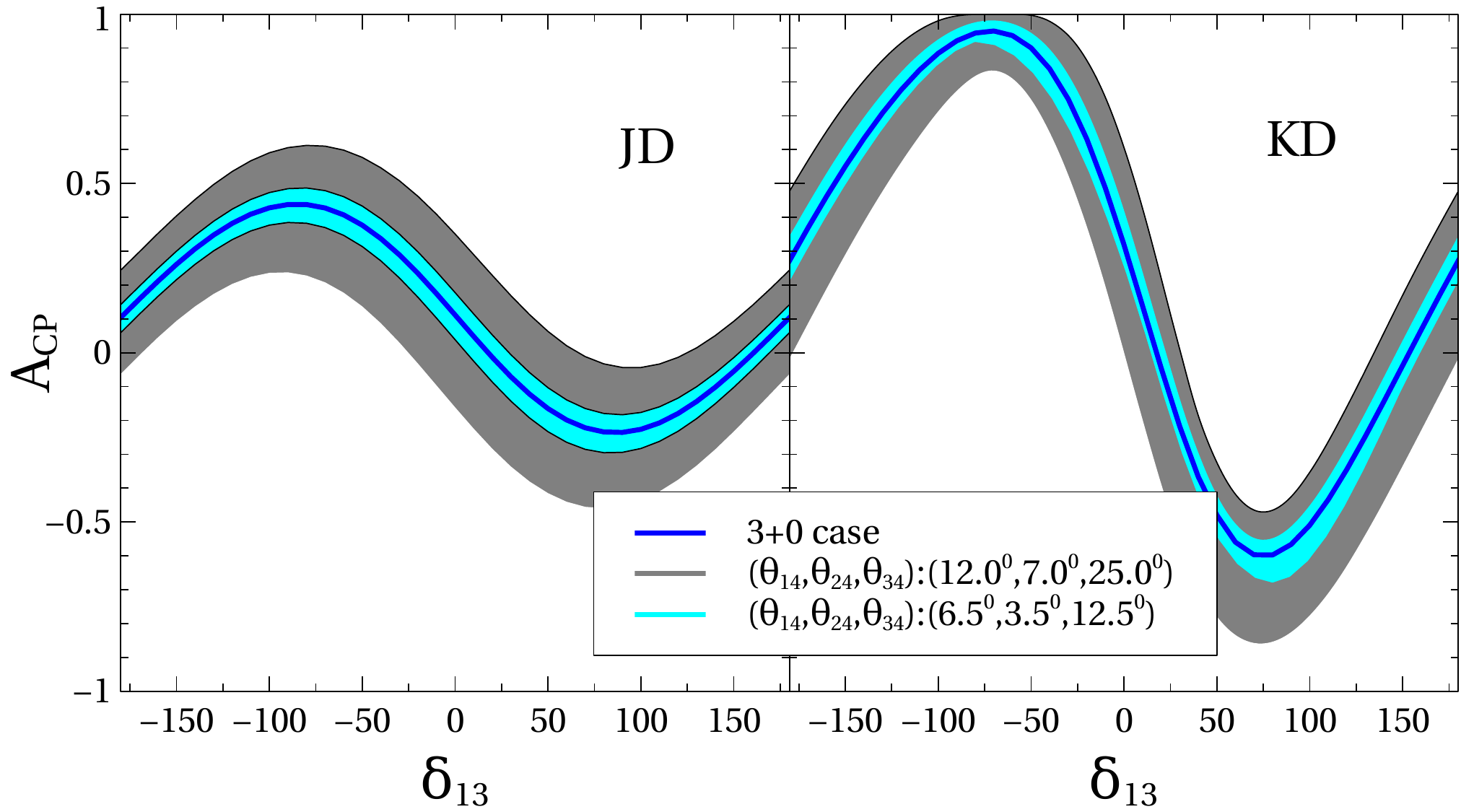}
\caption{CP-asymmetry at JD(left) and KD(right). The bands represent the variation of $\delta_{24}$ and $\delta_{34}$. }
\label{asym}
\end{figure}
Here in Fig.~\ref{asym}, we show the CP violation in terms of an asymmetry defined at the probability level to provide some additional insight. We define the CP asymmetry as 
\begin{equation}\label{def_asym}
A_{CP} = \frac{P_{\mu e}-\bar{P}_{\mu e}}{P_{\mu e}+\bar{P}_{\mu e}}\,.
\end{equation}
The left panel of Fig.~\ref{asym} 
shows the CP-asymmetry as a function of $\delta_{CP}$ for the JD baseline of 295 km and at a fixed energy 0.5 Gev which corresponds to the first oscillation maximum at this baseline. 
The right panel shows the CP-asymmetry for the KD baseline of 1100 km and at a fixed energy of 0.66 GeV  which corresponds to the second oscillation maximum for this baseline. The solid blue curves depict the $A_{CP}$ for the 3+0 $\nu$ case in both panels. The bands represent the effect of the sterile phases. For each $\delta_{13}$, the bands show the range of $A_{CP}$ for the full range of values of $\delta_{24}$ and $\delta_{34}$ between $[-\pi,\pi]$. The cyan bands show the effect of the sterile phases for the lower set of sterile mixing angles and the grey bands show the same for the higher set of mixing angles. We see that the presence of the sterile neutrino mixing spreads the $A_{CP}$ in both side of the standard 3+0 case. This implies that for a given true value of $\delta_{13}$, the CP asymmetry in the presence of sterile neutrinos could either increase or decrease for both JD and KD baselines and energies, depending on the true values of a $\delta_{24}$ and $\delta_{34}$. Looking from another angle, this also implies that presence of sterile neutrino mixing brings in an uncertainty in the expected CP asymmetry at long baseline experiments. Observation of a certain CP  asymmetry signal in the data could come from degenerate solutions involving mixing angles and 
phases in the 3+0 sector and the sterile sector. 
We can also see from the plots that with higher mixing, widths of the bands increase and hence the uncertainties in the CP-sector introduced by the sterile phases increases.
We also note that the CP-asymemtry is significantly higher for KD than for JD. 
The reason is that, the shape and magnitude of the curves largely depend on the baseline and energy value chosen, or more precisely
on the L/E factor.

\begin{figure}
\includegraphics[scale=0.75]{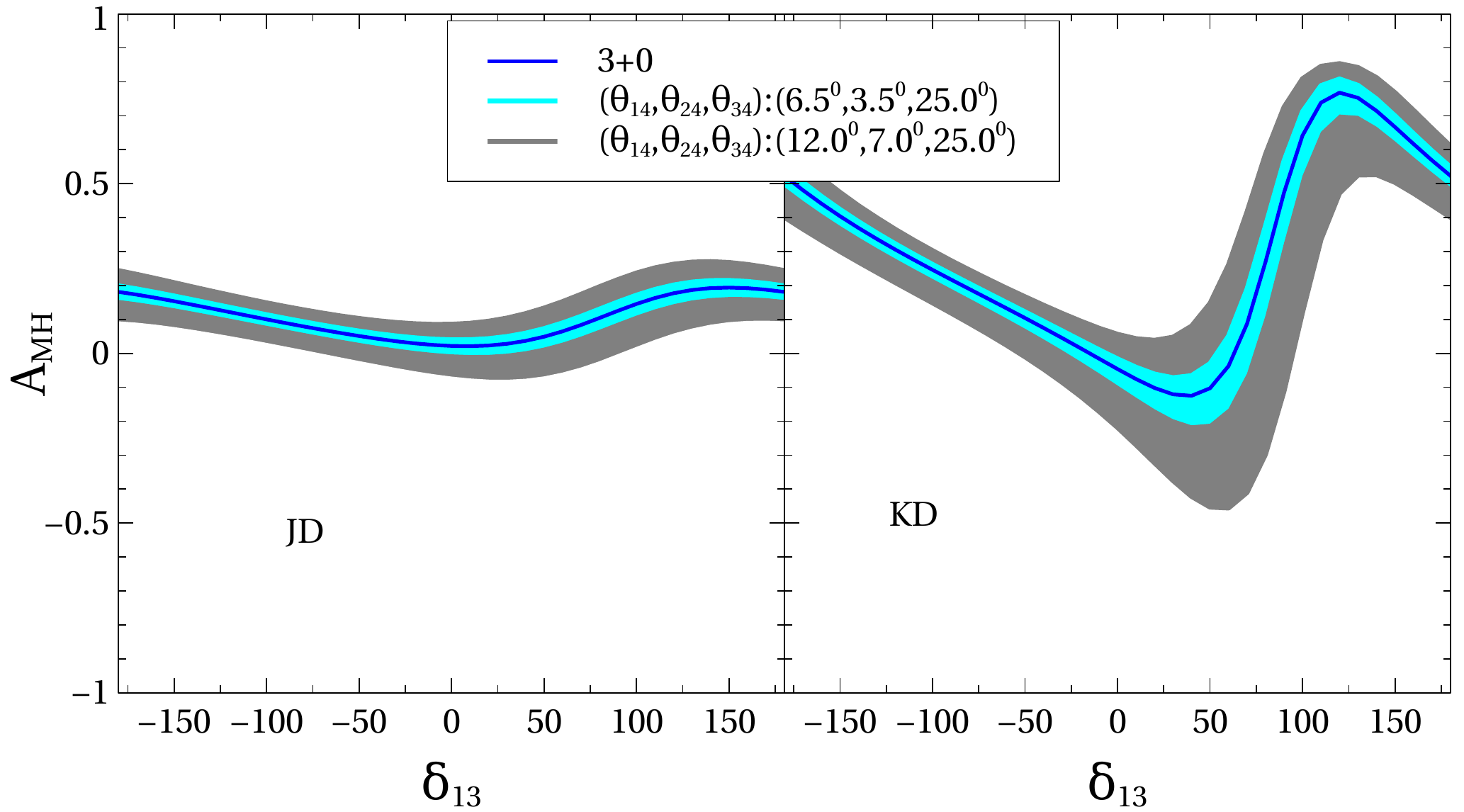}
\caption{MH-asymmetry as a function of $\delta_{13}$ for JD(left) and KD(right). The bands represent the variation of $\delta_{24}$ and $\delta_{34}$.}
\label{mhasym}
\end{figure}

The Fig.~\ref{mhasym} shows the MH-asymmetry for JD(left) and KD(right) baselines at a fixed energies of 0.5 GeV and 0.66 GeV, respectively. 
The MH-asymmetry is defined as:
\begin{equation}\label{def_mh}
A_{MH} = \frac{P^{NH}_{\mu e}-P^{IH}_{\mu e}}{P^{NH}_{\mu e}+P^{IH}_{\mu e}}
\end{equation} 

In NH we have taken $\ma = 2.45\times10^{-3}$ eV$^2$ while in IH we have taken $\ma = -\ma+\ms$. Just as in the previous figure, the blue solid curves show the MH-asymmetry for the standard (3+0) case. As before the bands are obtained when the sterile mixing angles $\delta_{24}$ and 
$\delta_{34}$ are varied in their full range $[-\pi,\pi]$. 
The cyan bands are for the lower set of sterile mixing angles and the grey bands are for higher set of sterile mixing angles. 
From the Fig.~\ref{mhasym}, we can see the effect of the new physics at the probability level. 
We can see naively that at the probability level it appears that for the 3+1 scenario, 
the mass hierarchy asymmetry has a chance of either increasing or decreasing compared to its
3+0 expected reach, depending on the true values of $\delta_{24}$ and $\delta_{34}$. 
We can see that with the increase of the sterile mixing angles the effect increases. 
As in Fig.~\ref{asym}, we find that the MH-asymmetry expected in KD is more than in JD and the 
reason for this is its higher chosen energy value and earth matter effects.

 \subsection{Statistical $\chi^2$}
 We define the $\chi^2$ (statistical) as follows: 
\begin{equation}
\chi^2(n^{true},n^{test},f) = 2\sum_{i}^{bins}\Big(n^{true}_{i}ln\frac{n^{true}_{i}}{n^{test}_{i}(f)}+n^{test}_{i}(f)-n^{true}_{i}\Big)+f^{2}
\end{equation}
, where $n^{true}_{i}$ and $n^{test}_{i}$ are the event rates that correspond to data and fit in the $i^{th}$ bin and f represents the nuisance parameters. For JD, KD and DUNE, we have 24, 10 and 39 bins respectively. Here in this definition, $n^{test}$ changes according to the sensitivity studied for a particular model. We have marginalised over the systematic uncertainties for each experiment as given in section $\rm IV$.
 
\subsection{CP sensitivity}
In this section, we present our CP violation sensitivity results in the presence of a light sterile neutrino. That is, we are addressing the following question: If CP is violated in Nature, then at what C.L., T2HK(JD$\times 2$) and T2HKK(JD+KD) can exclude the CP conserving scenarios in the presence of a sterile neutrino? In the 3+1 scenario, we have two more phases $\theta_{24}$ and $\theta_{34}$ in addition to the standard CP phase $\delta_{13}$. So while studying CP violation sensitivity of these experiments in the presence of sterile neutrinos, we consider the following two situations:
\begin{itemize}
\item[1.] CP is violated and we do not know the source of its violation. That is, it can be due to any of the three phases.
\item[2.] CP is violated and we know the source of its violation. For instance, say we assume that it is due to the standard Dirac CP phase $\delta_{13}$. 
\end{itemize} 
As explained above, we have presented our results for two sets of sterile mixing angles and for each set we fix the standard oscillation parameters to their best fit values in `data' and vary all the three true phases in their full range $[-\pi,\pi]$.
While answering the question of CP violation in the first scenario, we generate the data at a given true value of $\delta_{13}$, $\delta_{24}$ and $\delta_{34}$ and calculate the $\Delta\chi^{2}_{\rm min}$ by considering all the eight possible CP conserving scenarios in the `fit'. We marginalise over a fine grid of $\theta_{14}$, $\theta_{24}$, $\theta_{34}$ in their allowed range in the `fit'. In addition, we have marginalised over the test $\theta_{13}$ in its 3$\sigma$ allowed range.


\begin{figure}[h]
\includegraphics[scale=0.6]{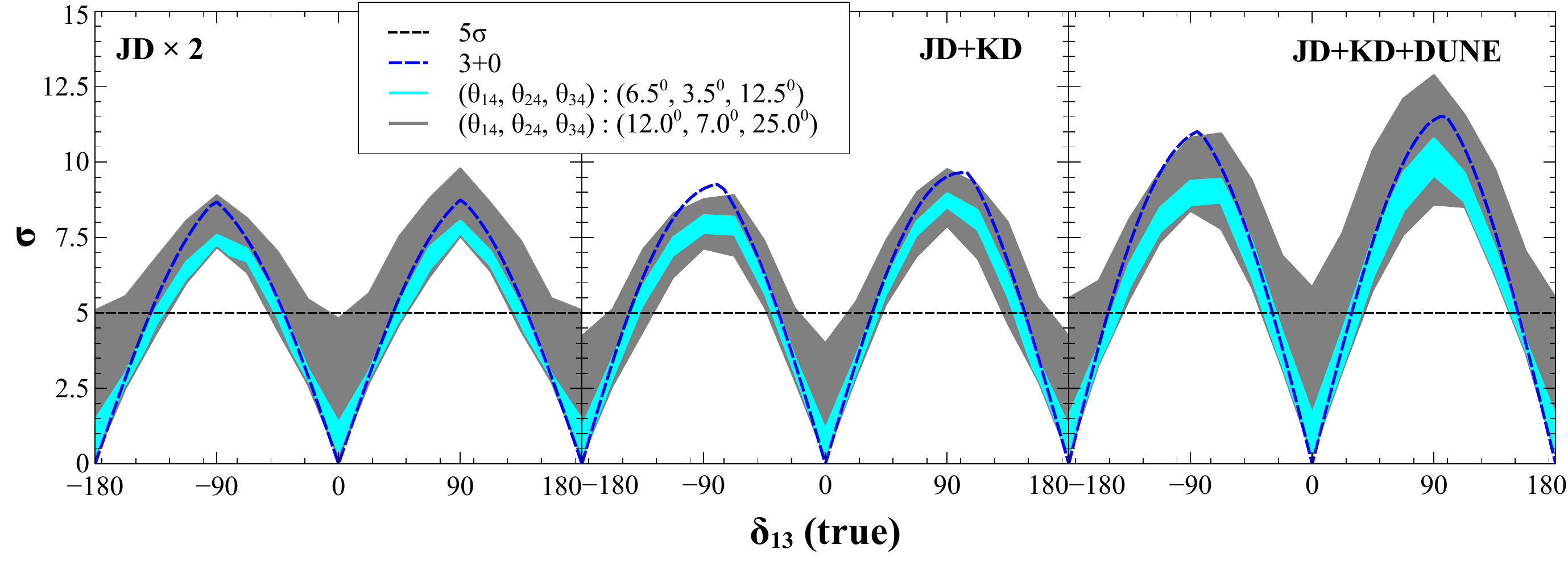}
\caption{The expected CP violation sensitivity of T2HK (JD$\times 2$), T2HKK (JD+KD) and DUNE+JD+KD under the assumption that we do not know the source of its violation. The bands correspond to variation of $\delta_{24}$ and $\delta_{34}$ in the true parameter space. The results are for true normal hierarchy.}
\label{cpv}
\end{figure}

The results of CP violation sensitivity in the first scenario are presented in Fig.~\ref{cpv}. The mass hierarchy is kept fixed as NH in both the `data' as well as the `fit' for this figure and in the next two figures in this subsection. The blue dashed line corresponds to the standard CP sensitivity in the 3+0 case as a function of $\delta_{13}$(true). The bands correspond to the 3+1 case with all possible choices for the other two phases $\delta_{24}$(true) and $\delta_{34}$(true). The thinner cyan band gives this band for $\delta_{24}$(true) and $\delta_{34}$(true) when the sterile mixing angles are taken to be at their smaller benchmark value while the grey band is obtained for the corresponding case when the sterile mixing angles are kept at their limiting benchmark value. In both cases these are the value of the sterile mixing angles in the `data', kept fixed in the entire band, while in the `fit' they are marginalised as discussed above. 
We observe that for all $\delta_{13}$(true) which give more than 5$\sigma$ CP sensitivity, the cyan band almost lies below the blue line. So for relatively small 3+1 mixing angles, sensitivity to CP violation in the 3+1 case decreases compared to the standard 3+0 case. But as the mixing angles increases, the 3+1 grey band spans both sides of the 3+0 plot. In the 3+1 case, as discussed in \cite{Dutta:2016glq}, there are two effects that come into play.  Firstly, the number of parameters to be marginalised in the `fit' is higher than the 3+0 case and marginalising over a large number of parameters brings the $\chi^2_{min}$ down. Secondly, effect of the variation of the true phases changes with the variation of the true values of the mixing angles. When mixing angles are small, effect of the true phases are also small and hence the first effect dominates the second one. As a result, the $\chi^2_{min}$ decreases compared to the 3+0 case. But as the true mixing angles increase, effect of the variation of the true phases increases simultaneously and as a result the width of the grey band increases. So for higher values of the mixing angles, the second effect tends to increase the $\chi^2_{min}$ and as a result overall sensitivity increases and the grey band spreads on both side of the 3+0 plot.  

As seen from Fig.~\ref{cpv}, T2HKK (JD+KD) has slightly better sensitivity to CP violation in the 3+0 case than T2HK(JD$\times 2$). But in the 3+1 case, we observe that T2HK and T2HKK have almost equal sensitivity to CP violation at $\delta_{13}=\pm \pi/2$ when mixing angle are large. But for smaller mixing angles, T2HKK gives slightly better sensitivity than T2HK (JD$\times 2$). Also note that in the 3+1 case, we have non zero CP violation sensitivity even at the CP conserving values of $\delta_{13}$(true). For the higher mixing angle case and for $\delta_{13}$(true)$=0$ or $\pm \pi$, we have more than $3\sigma$ CP violation sensitivity for $\delta_{24}$(true) in the range [$45\degree,135\degree$] and [$-135\degree,-45\degree$]. The effect of $\delta_{34}$(true) is less important. 
Combining JD and KD with DUNE enhances the CP violation sensitivity and we observe that even when $\delta_{13}$(true)$=0$ or $\pm \pi$,
it is possible to achieve more than 5$\sigma$ CP violation 
sensitivity in the higher mixing angle case for $\delta_{24}$(true) in the range  [$90\degree,112\degree$] and [$-112\degree,90\degree$]. Again the impact of $\delta_{34}$ is not important.

\begin{figure}[h]
\includegraphics[scale=0.6]{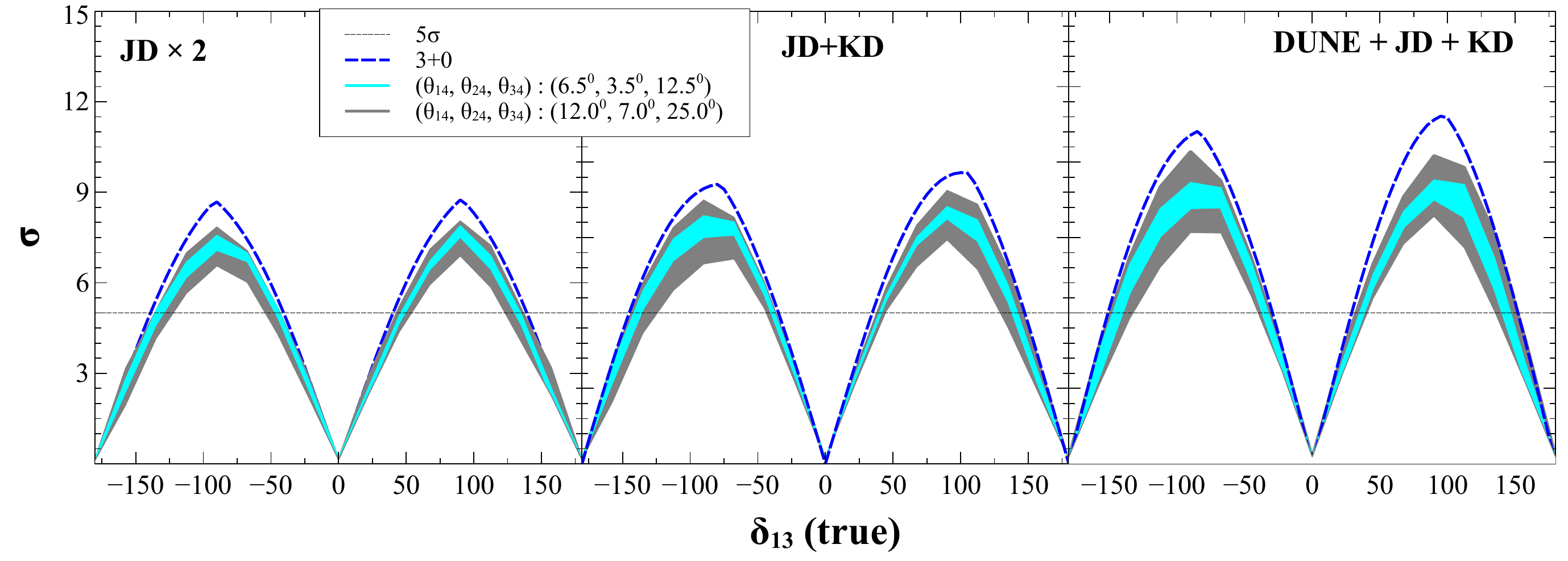}
\caption{The expected CP violation sensitivity of T2HK (JD$\times$ 2), T2HKK (JD+KD) and DUNE+JD+KD under the assumption that we know the source of its violation and it is due to $\delta_{13}$. The bands correspond to variation of $\delta_{24}$ and $\delta_{34}$ in the true parameter space. The results are for true normal hierarchy.}
\label{d13}
\end{figure}
\begin{figure}[h]
\includegraphics[scale=0.6]{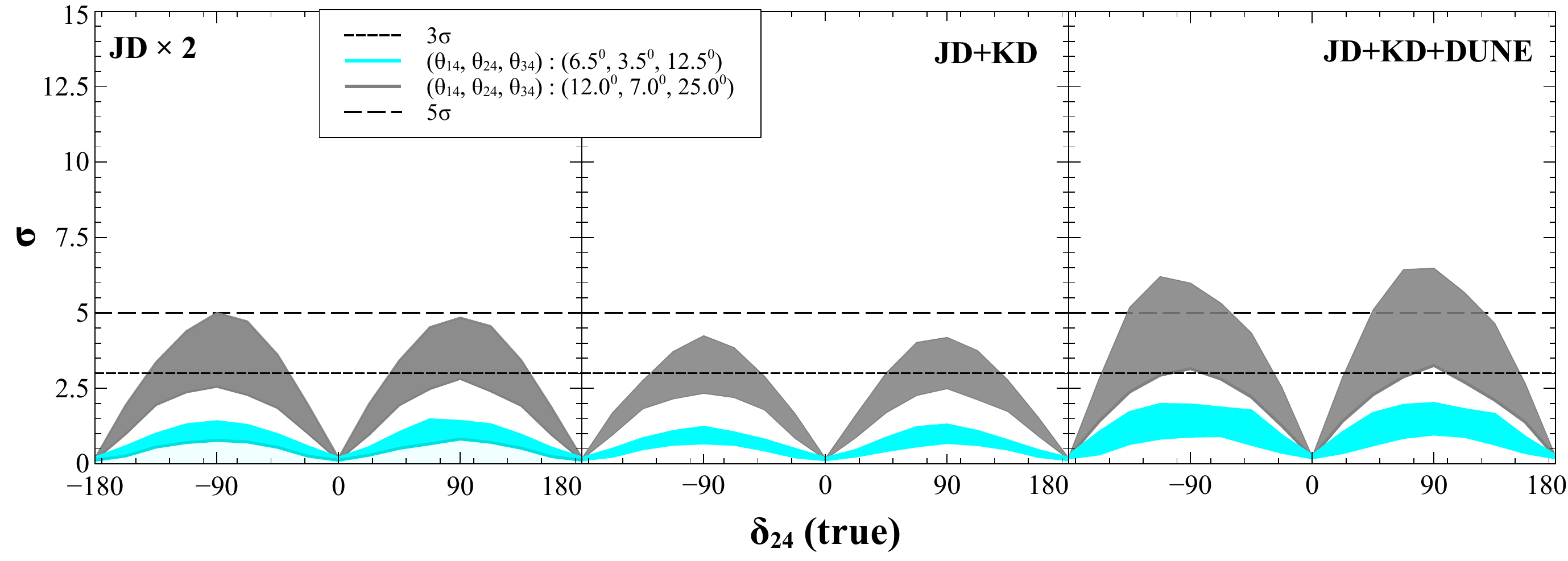}
\caption{The expected CP violation sensitivity of T2HK (JD$\times$ 2), T2HKK (JD+KD) and DUNE+JD+KD under the assumption that we know the source of its violation and it is due to $\delta_{24}$. The bands correspond to variation of $\delta_{13}$ and $\delta_{34}$ in the true parameter space. The results are for true normal hierarchy.}
\label{d24}
\end{figure}

To address the second scenario we consider the CP conserving cases for $\delta_{13}$ alone in the `fit' while we have marginalised over the two new phases $\delta_{24}$ and $\delta_{34}$ in the allowed 3$\sigma$ range. All other 3+1 mixing angles and $\theta_{13}$ are marginalised as explained above. The results are shown in Fig.~\ref{d13}, where we observe that the CP violation sensitivity in the 3+1 case decreases compared to the standard 3+0 sensitivity in all the three experimental set-ups. Both the cyan and grey bands lie below the 3+0 sensitivity plot for all true $\delta_{13}$. We also observe that the minima of the grey band is nearly at the same level in T2HKK and DUNE+T2HKK while it is slightly lower in the case of T2HK. From Fig.~\ref{d13} (Fig.~\ref{d24}), we observe that restricting the source of CP violation to only $\delta_{13}$ ( $\delta_{24}$) in the `fit' brings down the sensitivity . In the previous case (Fig.~\ref{cpv}), we have considered only eight combinations of the phases in the `fit' while in this case we have marginalised over two phases in their full allowed range. Marginalising over a large number of parameters lead to decrease in sensitivity. 

In Fig.~\ref{d24}, we have shown the results for CP violation due to $\delta_{24}$ and here the width of the bands are due to the variation of $\delta_{13}$(true) and $\delta_{34}$(true) in their full allowed range. Also in the `fit' we have marginalised over these two phases in their full range and choose CP conserving values for $\delta_{24}$. CP violation sensitivity in this case is much lower than the previous two cases and the two bands are well separated from each other. So if there is a sterile neutrino and CP violation is considered to be solely due to $\delta_{24}$, then T2HK and T2HKK can measure it at 3$\sigma$ C.L. only for some fraction of true $\delta_{24}$ and as seen from Fig.~\ref{d24}, T2HK is better than T2HKK. Even after combining with DUNE, we can achieve 5$\sigma$ CP violation sensitivity for some fraction of true $\delta_{24}$ only around the peak when the mixing angles are large. 

\subsubsection{CP Violation sensitivity if hierarchy is unknown}

The results presented in the previous section is under the assumption that the mass hierarchy will be known by the time of operation of these experiments. We have also checked the CP violation sensitivity for the true inverted hierarchy and the behaviours of the results are consistent.

\begin{figure}[h] 
\includegraphics[scale=0.6]{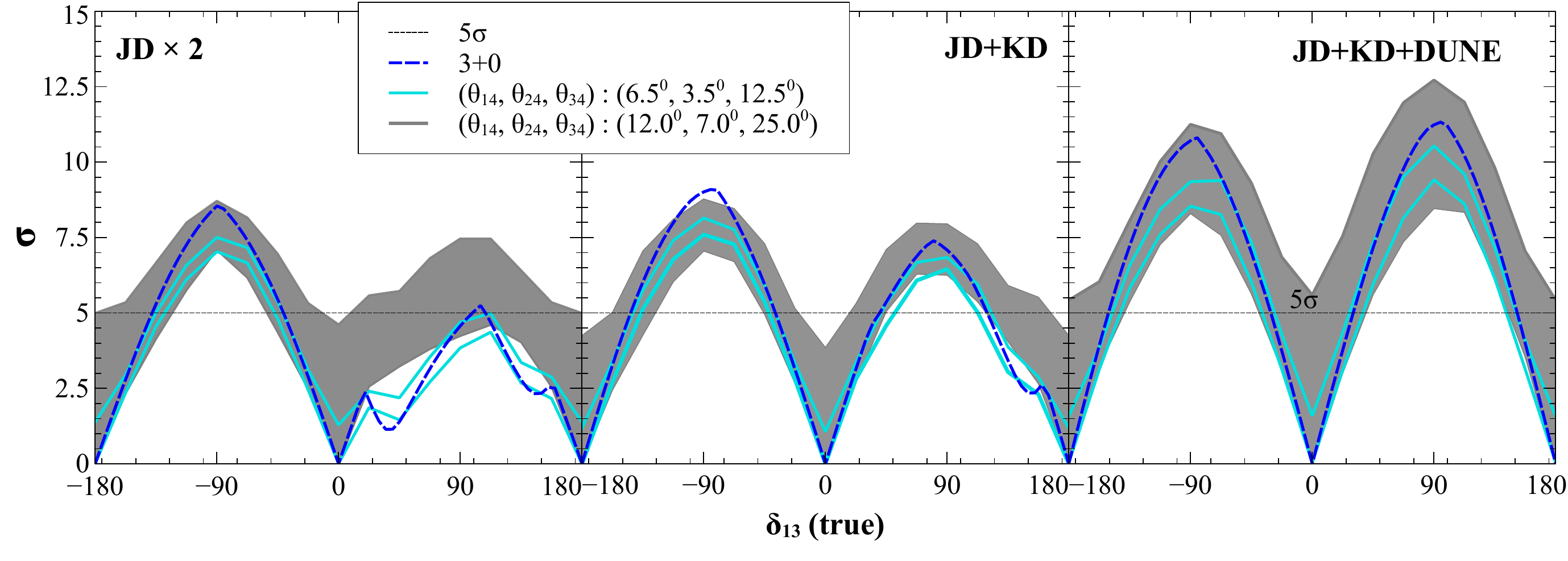}
\caption{The expected CP violation sensitivity of T2HK (JD$\times$ 2), T2HKK (JD+KD) and DUNE+JD+KD under the assumption that we do not know the source of its violation as well as the hierarchy. The bands correspond to variation of $\delta_{24}$ and $\delta_{34}$ in the true parameter space. The results are for true normal hierarchy.}
\label{cpvmh}
\end{figure}

In this section we show the CP violation sensitivity in the 3+1 case without fixing the mass hierarchy to its true case in the `fit'. In Fig.~\ref{cpvmh} we show the results for expected CP violation sensitivity for the first scenario where we rule out CP conserving values of all the three phases and with marginalisation over MH in the `fit'. We have assumed the NH to be true in this figure. That is, for each `simulated data' case we find the $\chi^2$ for eight combinations of CP conserving phases for each hierarchy in the `fit' and hence sixteen combinations in total. 
The minimum $\Delta\chi^2_{\rm min}$ amongst these sixteen combinations are plotted in Fig.~\ref{cpvmh} as a function of $\delta_{13}$(true) where the full range of $\delta_{24}$(true) and $\delta_{34}$(true) are represented  in the bands. 
Marginalisation over other parameters are same as explained in the previous subsection. The regions between the two cyan lines represent the CP violation sensitivity for the small mixing angles benchmark point, while the grey band depicts the sensitivity for the higher mixing angles benchmark case. The blue dashed lines show the 3+0 sensitivity for each of the three experimental set-ups. We observe that in the 3+0 scenario, CP violation sensitivity decreases significantly for T2HK, specially in the upper half plane (UHP) ($\delta_{13}$(true) $>0$) if the MH is taken to be unknown. For T2HKK although there is a drop in the sensitivity in the UHP for the 3+0 case compared to its counterpart in Fig.~\ref{cpv}, the effect of marginalising over hierarchy is less here than for T2HK.  
This happens because KD can resolve the mass hierarchy degeneracy to a large extent in the T2HKK set-up. Once we combine DUNE with T2HKK, the effect of the unknown hierarchy almost vanishes. On the other hand in presence of a sterile neutrino, the CP violation sensitivity decreases if the new mixing angles are small. But for the larger mixing angles case, the sensitivity increases in the UHP and for some combinations of  $\delta_{24}$(true) and $\delta_{34}$(true), we can have more than 5$\sigma$ sensitivity at $\delta_{13}=+\pi/2$ for T2HK. Adding DUNE not only enhances the sensitivity but also nullifies the effect of unknown MH even in the presence of a sterile neutrino.

\begin{figure}[t]
\includegraphics[scale=0.7]{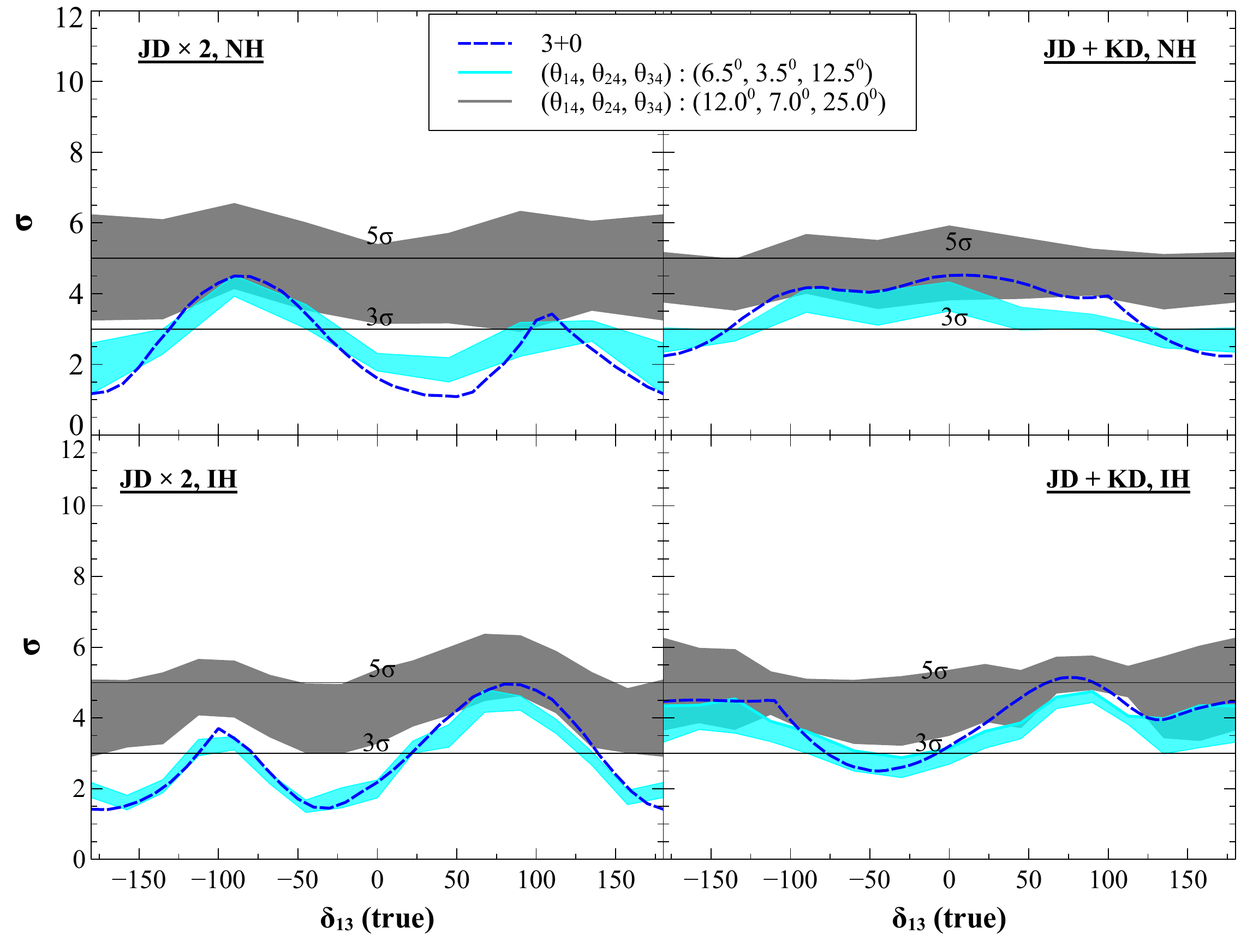}
\caption{The expected mass-hierarchy sensitivity of T2HK (JD$\times2$) and T2HKK (JD+KD). The upper panel is for true NH while the lower panel is for true IH. The bands correspond to the variation of the sterile phases in the true parameter space. }
\label{mh}
\end{figure}

\subsection{Mass hierarchy sensitivity}

In this section, we discuss the expected sensitivity to neutrino mass hierarchy of T2HK and T2HKK in the 3+1 scenario. For that we generate the data at a given true MH and `fit' it with the wrong hierarchy. We marginalise over all the three phases in the `fit' and all the sterile mixing angles and also $|\ma|$ and $\sin^2\theta_{23}$. Here we have not marginalised over $\theta_{13}$ due to computational challenges. We have seen in Fig.~\ref{mhasym} that the MH-asymmetry can change due to the presence of sterile neutrino mixing. However, there we have done everything only at the probability level and for fixed values of oscillation parameters as well as for a fixed energy. We will now show how the expected sensitivity to MH changes due to sterile neutrinos from a full analysis of expected data, when one takes all relevant marginalisation into account.  

The results for T2HK and T2HKK for both NH (upper panel) and IH (lower panel) as true are shown in Fig.~\ref{mh}. The presentation and description of the bands are same as that in the previous subsection. We observe that the sensitivity to mass hierarchy in presence of a sterile neutrino changes significantly in both the experiments. We have explicitly checked the effect of marginalisation on mass hierarchy sensitivity. Here also, we notice the two features discussed in section V B.  We observe that,

\begin{itemize}
\item[1.] For the higher sterile mixing angles, the impact of the variation of the true phases increases and it tends to increase the $\chi^2_{min}$.
\item[2.] Effect of marginalisation over test $|\Delta m^2_{31}|$ in the wrong hierarchy is very important for the MH sensitivity study. Its effect is seen to be most for the 3+0 case and keeps decreasing with the strength of the true sterile mixing. As a results the $\chi^2_{min}$ goes up for the higher set of sterile mixing angles.
\end{itemize}

For smaller sterile mixing angles, we observe that the width of the cyan band is small in both the hierarchies and the sensitivity for the 3+1 scenario in this case is nearly same as 3+0. In case of T2HK, the cyan band swings around the 3+0 plot but in T2HKK, it lies below the 3+0 line for most of the $\delta_{13}$(true) values, in both the hierarchies. However, the mass hierarchy sensitivity seems to increase for the larger mixing angle case. This apparently appears counter-intuitive to what we have observed in Fig.~\ref{biprob} where the overlap between the bi-probability plots was seen to increase when we increased the sterile mixing angles, making it appear as though the sensitivity to mass hierarchy would decrease as the sterile mixing angle was increased. We can also note that in Fig.~\ref{mh} the grey band does not span equally on both side of the cyan band. Instead the two bands get separated for most values of $\delta_{13}$(true). This again appears to be in contrast to Fig.~\ref{mhasym} where the cyan band was embedded nearly symmetrically inside the the grey band. The reason for both these apparent conflicts can be traced back to the impact of marginalisation of the mass hierarchy $\chi^2$ over $|\Delta m^2_{31}|$. We have checked that the marginalisation over $|\Delta m^2_{31}|$ alone reduces the mass hierarchy $\chi^2$ by nearly an order of magnitude for the 3+0 scenario. For the smaller sterile mixing angle case, the effect of sterile neutrino parameters is less and the final marginalised $\chi^2$ is close what we have for the 3+0 case. However, as the sterile mixing angle is increased the effect of marginalisation over 
$|\Delta m^2_{31}|$ is able to reduce the $\chi^2$ relatively less and as the result the final mass hierarchy sensitivity appears to rise. For the same reason the cyan band becomes asymmetric with respect to the grey band. For the largest sterile mixing angle case we have more than 5$\sigma$ sensitivity for mass hierarchy for any $\delta_{13}$(true).

\begin{figure}[h] 
\centering
\includegraphics[scale=0.7]{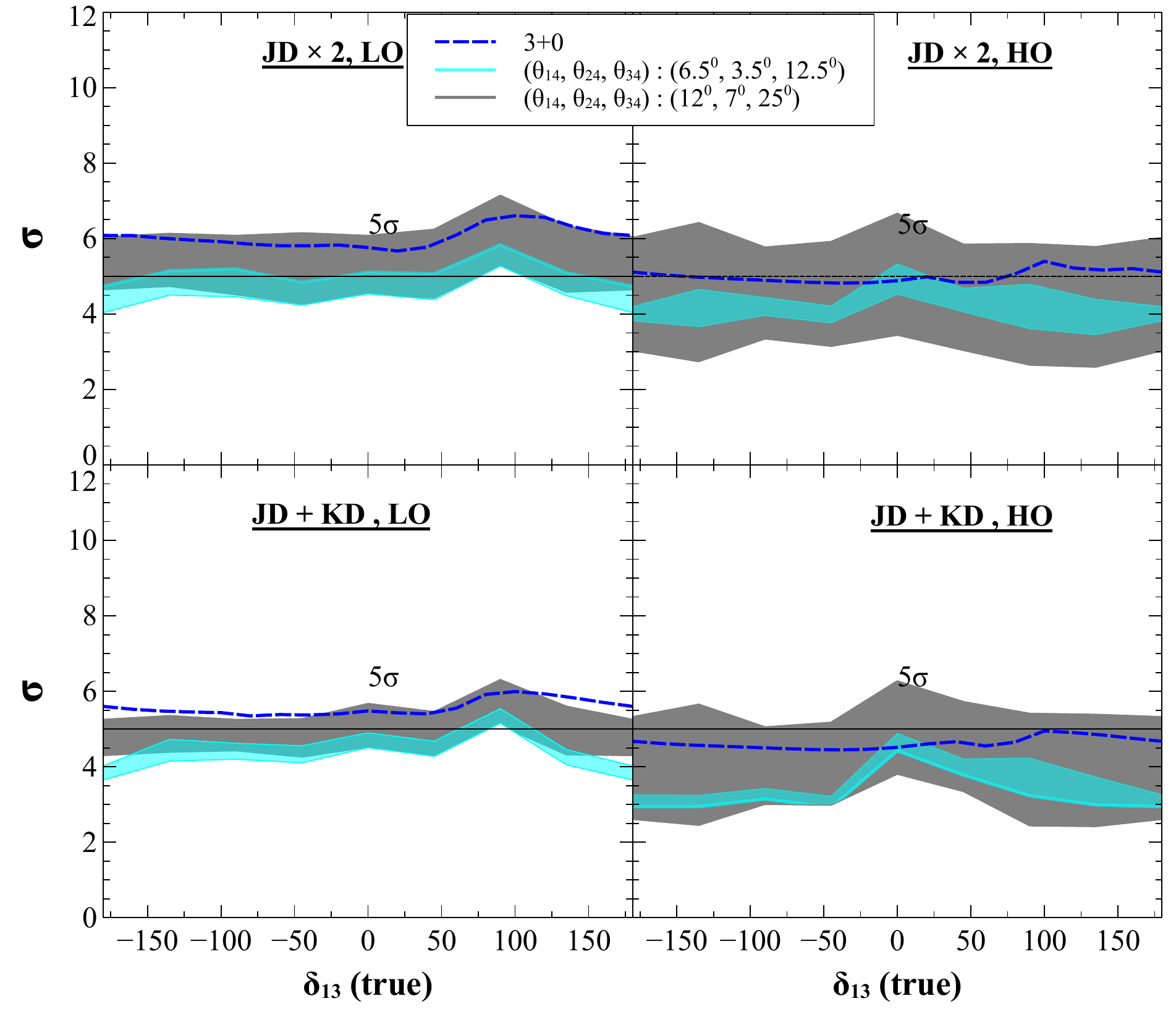}
\caption{\footnotesize{The expected octant discovery potential of T2HK (JD$\times$2) and T2HKK (JD+KD). In the LO (HO), we consider $\theta_{23}$ = 40.3$^{o}$ (49.7$^{o}$) as the true value. The upper panel is for T2HK (JD$\times$2) while the lower panel is for T2HKK (JD+KD). The bands correspond to the variation of the sterile phases.}}
\label{oct}
\end{figure}
\begin{figure}[h] 
\centering
\includegraphics[scale=0.7]{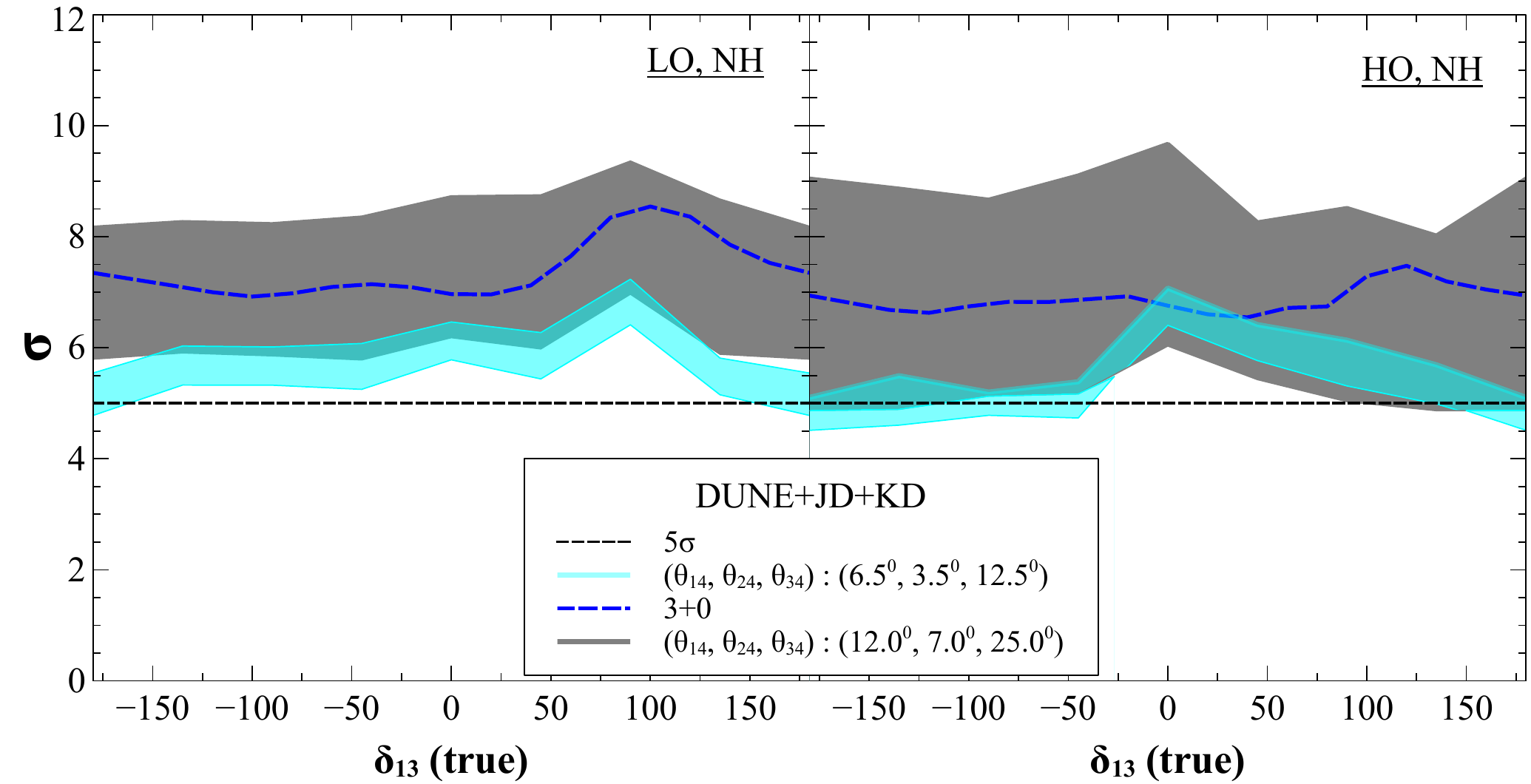}
\caption{The octant discovery potential of DUNE+T2HKK(JD+KD).}
\label{oct1}
\end{figure}

\subsection{Octant Discovery}

In this section, we study the octant discovery potential of T2HK and T2HKK in the 3+1 scenario. We also present a combined analysis of T2HKK with DUNE. We consider the same benchmark value of true $\theta_{23}$ for which we have shown the bi-probability plots in Fig.~\ref{biprob}. We then show the potential of these experiments to exclude the wrong octant for the two sets of sterile mixing angles. In data, we vary all the three phases, as discussed above, we show the band of $\Delta\chi^2_{\rm min}$ as a function of the $\delta_{13}$(true). The band corresponds to the full range of $\delta_{24}$(true) and $\delta_{34}$(true). 
In the `fit', we marginalise over the sterile mixing angles and the phases in their allowed ranges. In addition, we marginalise over $\theta_{13}$ and for a true LO (HO), $\theta_{23}$ is marginalised in the HO (LO). We consider the full 3$\sigma$ allowed range for both $\ty$ and $\tz$. We show the results only for the assumed true normal mass hierarchy. 

From Fig.~\ref{oct}, we observe that the potential of T2HK to exclude the wrong octant is slightly higher than for T2HKK. If LO is the true octant in the standard 3$\nu$ scenario, then both T2HK and T2HKK can exclude the wrong HO at 5$\sigma$ for all $\delta_{13}$(true). However for true HO, excluding the LO is not possible at 5$\sigma$ for all $\delta_{13}$(true), though the expected sensitivity reach is seen to remain high. In the presence of a sterile neutrino, the expected sensitivity gets modified and we observe that for small sterile mixing, the potential to exclude the wrong HO is significantly less than the standard 3+0 case. For large mixing angles, the width of the band increases and for some combinations of $\delta_{24}$(true) and $\delta_{34}$(true), the grey band slightly crosses the 3+0 plot for some fraction of $\delta_{13}$(true). On the other hand for true HO, the standard 3+0 plot lies within the grey band while the cyan band lies below it except for some small fraction of $\delta_{13}$(true) around zero. Also, one can note that for large sterile mixing angles, some combinations of the new phases give more than 5$\sigma$ HO discovery potential for both the set up.

If we add DUNE with T2HKK (Fig.~\ref{oct1}), the octant discovery potential changes significantly. For large (small) sterile mixing angles, if LO is the true octant, then it is possible to exclude the HO at 5$\sigma$ for all $\delta_{13}$(true) (all $\delta_{13}$(true) except some fraction around $\delta_{13}$(true)$=\pm\pi$), irrespective of the true values of the other two phases. Even in the case of true HO, we can rule out the LO at 5$\sigma$ C.L. except for some fraction of $\delta_{13}$(true).

\section{Summary \& Conclusion}

The final word on existence or nonexistence of sterile neutrino oscillations is yet to be affirmatively answered. The LSND 
hint for $\bar\nu_\mu \to \bar\nu_e$ oscillations as well as the reactor and gallium 
experiment anomalies demand that there be an extra light neutrino state which must be sterile 
with $\Delta m^2 \sim$ 1 eV$^2$ and mixed the active neutrino, albeit weakly. The presence of the mixed sterile 
state brings in additional mixing angles and phases that affect the oscillation probabilities at long baseline experiments. 
In particular, while the additional mass squared  difference drops out due to averaging of the fast oscillations at the 
long baselines, all the three active-sterile mixing angles $\theta_{14}$, $\theta_{24}$ and $\theta_{34}$ and the two 
additional phases $\delta_{24}$ and $\delta_{34}$ impact the neutrino oscillation probabilities. 
The active-sterile mixing angles $\theta_{14}$, $\theta_{24}$ and $\theta_{34}$, 
even though constrained by data from Daya Bay, IceCube and MINOS, respectively, bring in additional 
uncertainty in the oscillation probabilities, while the new phases $\delta_{24}$ and $\delta_{34}$ are totally 
unconstrained. Hence the CP violation sensitivity is obviously affected, with the possibility of additional CP violation 
coming from the new phases $\delta_{24}$ and $\delta_{34}$. The mass hierarchy sensitivity and the octant 
sensitivity also gets affected.

In this work we studied the impact of this active-sterile mixing angles and new phases on the physics reach of the 
T2HK and T2HKK proposals. We also studied the combined sensitivity reach of T2HKK and DUNE. We showed 
the impact of the sterile mixing angles and phases on T2HK  and T2HKK 
using the bi-probability plots. We further illustrated the expected CP violation sensitivity for T2HK and T2HKK 
by plotting the the CP asymmetry as well as the mass hierarchy asymmetry 
as a function of the standard CP phase $\delta_{13}$ while the sterile phases 
$\delta_{24}$ and $\delta_{34}$ were allowed to vary in the full range. In all cases we showed the corresponding 
line for the standard 3+0 case in order to illustrate the change brought by the 3+1 scenario. As expected these 
plots showed that the spread in the sensitivity due to the sterile sector increased as the sterile mixing angles increased. 

We showed the expected CP violation sensitivity of T2HK and T2HKK, and the combined CP reach of T2HKK+DUNE as a function of 
$\delta_{13}$(true), while $\delta_{24}$(true) and $\delta_{34}$(true) were allowed to take any value in their full range 
[$-\pi,\pi$] and the corresponding band was plotted. 
This study was performed first by checking against CP conserving cases for all three phases 
$\delta_{13}$, $\delta_{24}$ and $\delta_{34}$ in the `fit' and then by checking only against the 
CP conserving cases for $\delta_{13}$ and marginalising over the others. It was shown that while in the latter case it might appear 
that the CP violation sensitivity of the experiment is going down, in the former more general case the 
sensitivity might even increase over that expected for the 3+0 case for some value of $\delta_{24}$(true) and $\delta_{34}$(true).
The combined T2HKK+DUNE data could give us a CP violation sensitivity of more than 12.5$\sigma$ for 
certain values of $\delta_{24}$(true) and $\delta_{34}$(true) and $\delta_{13}$(true) $=\pi/2$. 
Also, even for $\delta_{13}$(true) $=0^\circ$ we can get a greater than $5\sigma$ sensitivity for CP violation 
from the combined data from T2HKK and DUNE. We showed also the CP violation sensitivity in these experiments 
as a function of $\delta_{24}$(true) when $\delta_{13}$(true) and $\delta_{34}$(true) were allowed to take all possible 
values and the corresponding sensitivity was plotted as a band. Finally, we showed the CP violation sensitivity by 
marginalising over the hierarchy. It was seen that the CP sensitivity of T2HK for $\delta_{13}$(true) $=\pi/2$ goes 
down. However, for T2HKK and T2HKK+DUNE the sensitivity remains mostly unchanged mainly due to the fact that 
these set-ups are able to resolve the issue of hierarchy themselves owing to their longer baselines. 

We also showed how the expected mass hierarchy sensitivity of 
T2HK, T2HKK and T2HKK+DUNE changes in the presence of sterile mixing. Here the effect of increasing 
the value of the sterile mixing angle was shown to have an interesting effect on the corresponding expected sensitivity of the 
T2HK and T2HKK set-ups. When the sterile mixing angles are taken to be small, 
the neutrino mass hierarchy sensitivity of T2HK and T2HKK is seen to deteriorate 
in the 3+1 case compared to the 3+0 case, 
for some values of $\delta_{13}$(true), $\delta_{24}$(true) and $\delta_{34}$(true). 
However, when the sterile mixing angles were increased the mass hierarchy sensitivity 
of T2HK and T2HKK increased for nearly all values of 
$\delta_{13}$(true), $\delta_{24}$(true) and $\delta_{34}$(true). 
The reason for this behavior was traced to the effect of marginalisation over $|\Delta m^2_{31}|$. 

Finally, we presented the expected octant of $\theta_{23}$ sensitivity of T2HK, T2HKK and T2HKK+DUNE 
in the 3+1 scenario. We showed at results for both choices of LO and HO and for two sets of 
true sterile mixing values and the full range of all the three phases 
$\delta_{13}$(true), $\delta_{24}$(true) and $\delta_{34}$(true). Again for this case, the 
expected octant sensitivity appears to be mostly decreasing in the 3+1 case compared to the 3+0 
case when the mixing angles are small. But when the true sterile mixing angles are taken to be 
larger, the octant sensitivity is seen to improve for some sets of values of 
$\delta_{13}$(true), $\delta_{24}$(true) and $\delta_{34}$(true), especially for T2HKK and 
T2HKK combined with DUNE.

\section*{Acknowledgment}
We acknowledge the HRI cluster
computing facility (http://cluster.hri.res.in).
The authors would like to thank the Department of Atomic Energy
(DAE) Neutrino Project under the XII plan of Harish-Chandra
Research Institute.
This project has received funding from the European Union's Horizon
2020 research and innovation programme InvisiblesPlus RISE
under the Marie Sklodowska-Curie
grant  agreement  No  690575. This  project  has
received  funding  from  the  European
Union's Horizon  2020  research  and  innovation
programme  Elusives  ITN  under  the 
Marie  Sklodowska-
Curie grant agreement No 674896.

\bibliography{ref}

\end{document}

%% file: header.tex
\newcommand{\be}{\begin{eqnarray}}
\newcommand{\ee}{\end{eqnarray}}

\def\nue{{\nu_e}}

\def\numu{{\nu_{\mu}}}

\newcommand{\ms}{\Delta m^2_{21}}
\newcommand{\ma}{\Delta m^2_{31}}

\def\gs{\mathrel{
   \rlap{\raise 0.511ex \hbox{$>$}}{\lower 0.511ex \hbox{$\sim$}}}}
\def\ls{\mathrel{
   \rlap{\raise 0.511ex \hbox{$<$}}{\lower 0.511ex \hbox{$\sim$}}}}
\newcommand{\bea}{\begin{equation} \begin{array}{c}}
\newcommand{\bead}{\begin{equation} \begin{array}{cccc}}
\newcommand{\eea}{ \end{array} \end{equation}}